\documentclass[twocolumn,secnumarabic,amsmath,amssymb,nobibnotes, aps, prl,superscriptaddress,]{revtex4-2}
\usepackage{graphicx}
\usepackage{dcolumn}
\usepackage{siunitx}
\usepackage{multirow}
\usepackage{color}
\usepackage{threeparttable}

\usepackage[mathlines]{lineno}

\begin{document}

\title{Anomalous nuclear effects on ion charge state distribution in helium gas}

\author{S.~Kimura}
\email[]{sota.kimura@kek.jp}
\affiliation{Wako Nuclear Science Center (WNSC), Institute of Particle and Nuclear Studies (IPNS), High Energy Accelerator Research Organization (KEK), Wako 351-0198, Japan}

\author{M.~Wada}
\altaffiliation[Present address: ]{Institute of Modern Physics, Huizhou branch, Chinese Academy of Science, Huizhou 516000, China
}
\affiliation{Wako Nuclear Science Center (WNSC), Institute of Particle and Nuclear Studies (IPNS), High Energy Accelerator Research Organization (KEK), Wako 351-0198, Japan}

\author{H.~Haba}
\affiliation{RIKEN Nishina Center for Accelerator-Based Science, Wako 351-0198, Japan}

\author{H.~Ishiyama}
\affiliation{RIKEN Nishina Center for Accelerator-Based Science, Wako 351-0198, Japan}

\author{T.~Niwase}
\affiliation{Department of Physics, Kyushu University, Fukuoka, 819-0395, Japan}

\author{M.~Rosenbusch}
\affiliation{RIKEN Nishina Center for Accelerator-Based Science, Wako 351-0198, Japan}

\author{P.~Schury}
\affiliation{Wako Nuclear Science Center (WNSC), Institute of Particle and Nuclear Studies (IPNS), High Energy Accelerator Research Organization (KEK), Wako 351-0198, Japan}
\date{\today}

\begin{abstract}
The influence of isotope differences on ion charge state yield ratios has never been studied in detail, having been considered negligible. However, we have observed anomalous ion charge state distributions in the thermalization of energetic atomic ions in helium gas; the charge state distributions varied between not only isotopes but also between nuclear states within the same nuclide. The magnitude of the observed results suggests that this anomaly is a universal phenomenon that cannot be explained by the framework of the known isotope effects. Nuclear spin and deformation could be key to unraveling this, but the mechanisms remain an open question. 
\end{abstract}

\maketitle


The charge states of the ions moving in materials reflect the relationship between their properties. Since the first work successfully observed the charge change of $\alpha$-particles by Henderson in 1922 \citep{Henderson1922}, many experimental studies of the ion charge state distributions (ICSDs) resulting from the interactions with materials were performed \citep{Betz1972, Wittkower1973, Shima1986, Shima1992, Leon1998}. In nuclear physics, the information on ICSDs is important for evaluating ion energy loss and range, design of charge strippers, operating gas-filled ion separators, etc. Many (semi-)empirical formulas estimating the ICSDs have been proposed for these purposes \citep{Ghiorso1988, Shima1989, Oganessian1991, Scheidenberger1998, Schiwietz2001, Oganessian2001, Liu2003, Kuboki2010, Kaji2011, Gregorich2013, Khuyagbaatar2013}. 

For slow ions ($v < {\rm Bohr~velocity}$) in a dilute gas, the ICSDs have been discussed solely based on the relative ionization potentials of the slow ions and the gas molecules \citep{Kudryavtsev2008, Schury2017, Kaleja2020, Mollaebrahimi2023}.  In the higher-velocity regime, the dominant electron transfer (ET) mechanism is the instantaneous process described by the velocity matching between the gas molecule's valence electrons and unoccupied orbits in the ion. The empirical formulas describing this process are constructed based on a semiclassical notion that the mean charge should be roughly proportional to the cube root of the ion's atomic number. The ICSD's elemental dependence has been at the center of interest in both velocity regions so far. 

In contrast with this, the isotopic dependence of ET has been considered negligible except in limited cases, for example, the ETs of ${\rm He}^{2+}$ ions with the hydrogen isotopes of H, D, and T \citep{Stolterfoht2007}. The most salient property difference among isotopes (or nuclear states of a given isotope) is their nuclear spin. The effect of nuclear spin on atomic states, known as the hyperfine interaction, has a typical magnitude of energy shift in atomic states due to hyperfine splitting of $\sim$$10^{-5}$~eV. In contrast, the energy scale of the ET between atoms is on the order of a few electron volts. Therefore, the influence of nuclear spin on ET had been presumed to be insufficient to meaningfully affect the ICSD.  However, during a systematic study of radioactive ions from a fission source stopped in a helium gas cell \cite{Kimura2024}, we discovered an anomaly: the existence of large isotope and nuclear state dependencies in the ICSDs.


The experimental setup used to perform these measurements is described in detail elsewhere \citep{Kimura2024}. The various radioactive ions produced via spontaneous fission of a 9.25-MBq ${}^{252}{\rm Cf}$ source ($T_{1/2}=2.645~{\rm years}$, fission branch:~3.09\%) were collected by a cryogenic helium gas cell (CHeGC). After the ions pass through the thin Mylar window sealing the CHeGC, their charge states are expected to be at least $q = 15+$ (hereafter, specific charge states will be denoted as n+). In the course of thermalizing in the CHeGC, they experience ETs with the cryogenic helium gas, and their median charge state reduces to 2+. The thermalized ions were quickly ($T\ll100$~ms) extracted by a radio-frequency carpet \citep{Wada2003} without contacting the component parts of the CHeGC and then were transported to a flat trap, the injector for a high-resolution multi-reflection time-of-flight mass spectrograph (MRTOF-MS) \citep{Schury2014}. Their times-of-flight (TOF), defined as the duration between ejection from the flat trap and implantation on the ion detector, were measured with the MRTOF-MS. The CHeGC was cryogenically cooled throughout the measurements, and the filling helium gas was provided from helium cylinders of purity $> 99.99995~{\rm vol}\%$ which was further purified with a three-stage purification system consisting of two active getters (SAES MonoTorr and MicroTorr) to remove hydrocarbons and other chemically active contaminants and a sub-10~K cold trap to suppress noble gas contaminants. This purification system facilitates the removal of contaminants to achieve an extremely clean environment inside the CHeGC.

\begin{figure*}[t]
	\begin{center}
		\includegraphics[width=0.85\textwidth, bb=0 0 425 340, clip, trim=0 0 0 0]{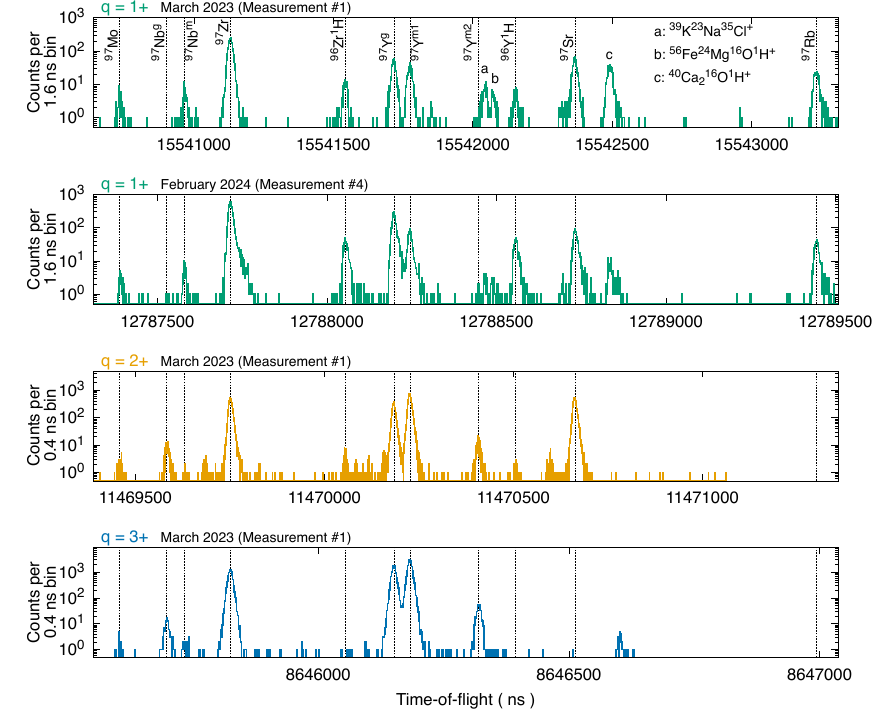}
		\caption{\label{Hist_TOF}Time-of-flight spectra focusing on $A = 97$ isobar series. The vertical dotted lines show the predicted position of each ion species. The unlabeled peaks in $q=1+$ TOF spectra are the molecules of stable isotopes. For $q = 1+$, the results of the measurements at two different periods are shown. }
	\end{center}
\end{figure*}

\begin{figure*}[t]
	\begin{center}
		\includegraphics[width=0.85\textwidth, bb=0 0 425 425, clip, trim=0 0 0 0]{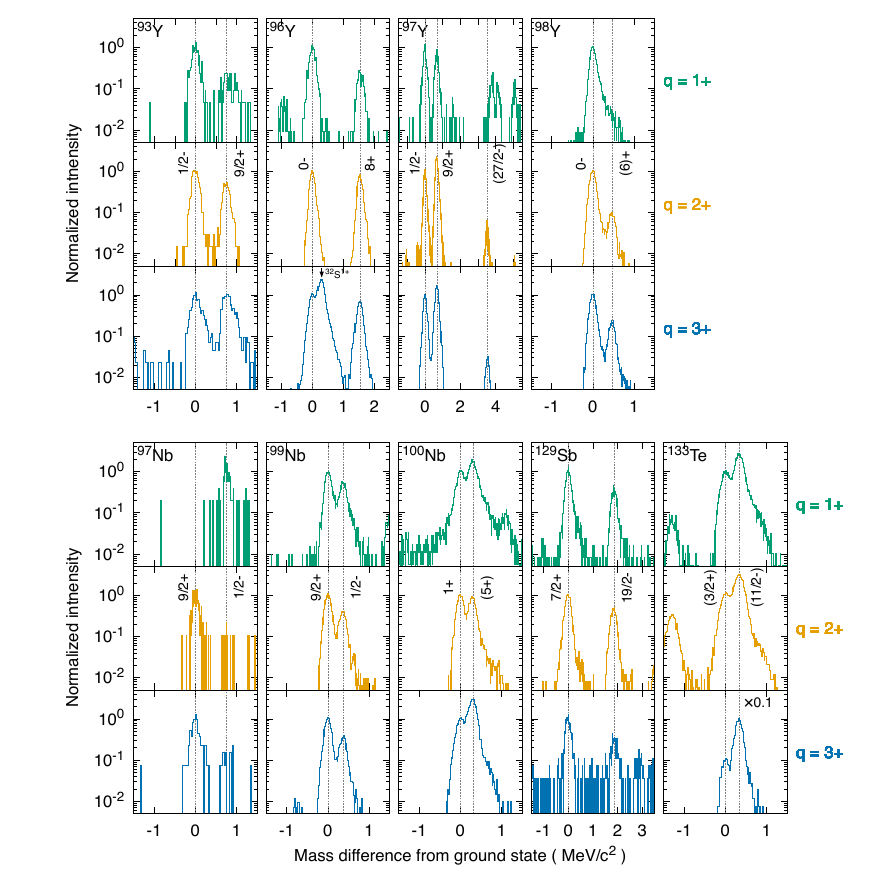}
		\caption{\label{Hist}Time-of-flight spectra where the horizontal axis has been converted to the mass-energy relative to the ground state. Peak intensities are normalized to the ground state intensity of each nuclide. The dashed lines show the position of the ground and the isomeric states of interest; 97Y has two isomeric states. The spin-parity of each state is also indicated with the dashed line; the parentheses denote prediction values.}
	\end{center}
\end{figure*}

\begin{figure*}[t]
	\begin{center}
	        \includegraphics[width=0.9\textwidth, bb=0 0 1020 283, clip, trim=0 0 0 0]{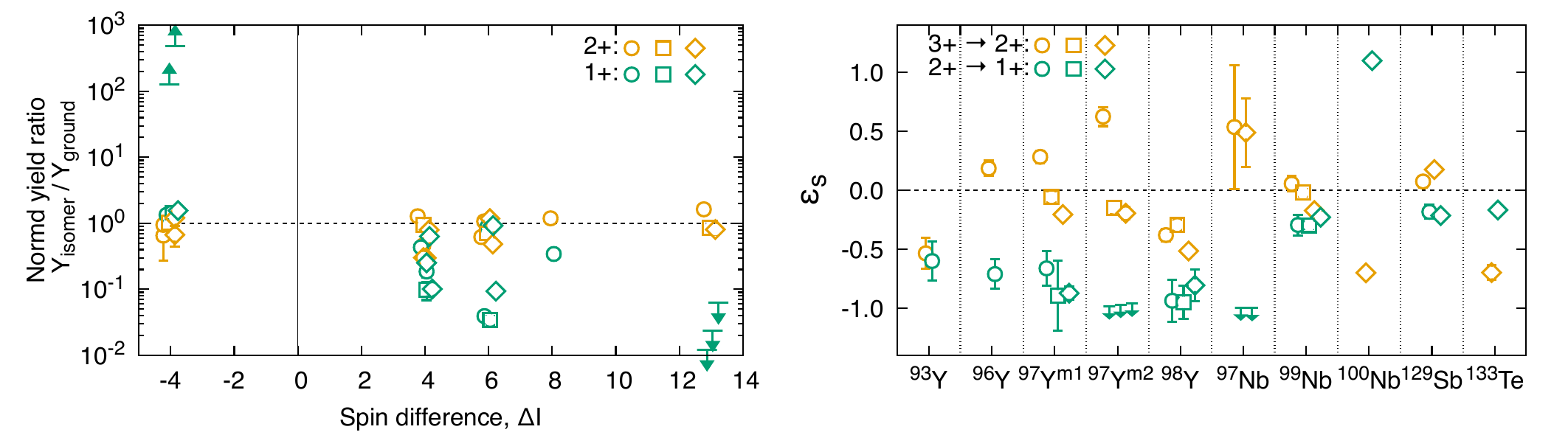}
		\caption{(Left panel) Normalized yield ratios of the ground and isomeric state's peaks as the function of nuclear spin difference: $\Delta I = I_{\rm isomer} - I_{\rm ground}$. All values are normalized by the values of the 3+ state. Several series of measurements were performed; the result of each is shown with a unique symbol shape. The arrows indicate the upper or lower limits. (Right panel) Spin enrichment factor (see Eq.~\ref{DefineOfES}), $\varepsilon_{\rm S}$, regarding the ground and isomeric states pair.\label{esOfIsomer}}
	\end{center}
\end{figure*}

\begin{figure*}[t]
	\begin{center}
		\includegraphics[width=0.9\textwidth, bb=0 0 510 255, clip, trim=0 0 0 0]{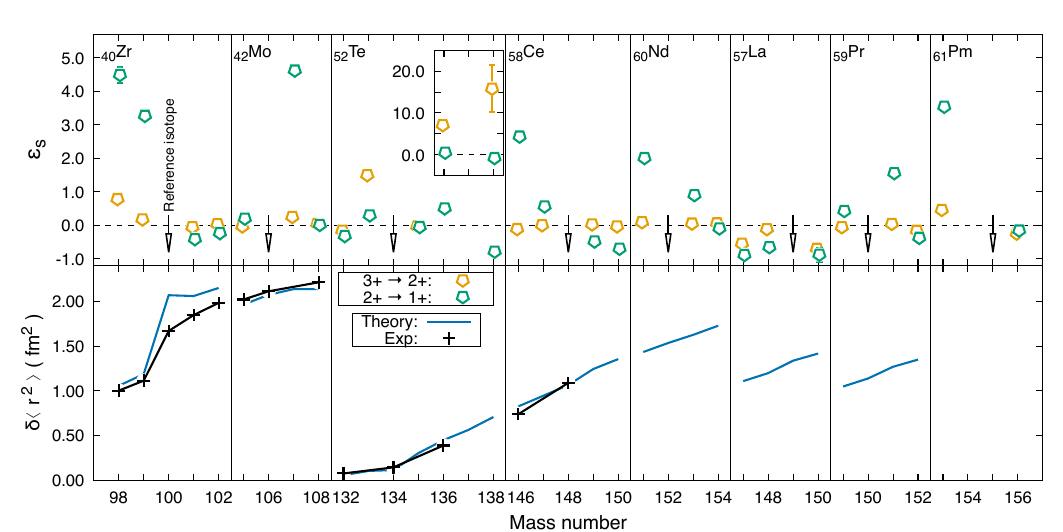}
		\caption{Spin enrichment factor (see Eq.~\ref{DefineOfES}), $\varepsilon_{\rm S}$, regarding the pairs of isotopes and change of mean square nuclear charge radii, $\delta \langle r^2 \rangle$. The arrows indicate the reference isotopes corresponding to the lower-spin component in Eq. 2. The other notations in the upper plot are the same as the right panel of Fig.~\ref{esOfIsomer}. The evaluated experimental data and the theoretical prediction of $\delta \langle r^2 \rangle$ are taken from \citep{Angeli2013} and \citep{Li2023}, respectively. \label{esOfIsotope}}
	\end{center}
\end{figure*}

Figure~\ref{Hist_TOF} shows a set of TOF spectra focusing on the mass number $A$~=~97 isobar series as an example of the fission fragment measurements with the MRTOF-MS. Both atomic and molecular ions can be found in the TOF spectrum. The intensity ratios of the observed ions are not constant across charge states due to differences in the chemical properties of elements; for example, the peak of ${}^{97}{\rm Rb}$ ( ${}^{97}{\rm Sr}$) can be found in the spectrum of 1+ (2+) but is not observed in that of 2+ (3+).

As the ET process varies negligibly between nuclear ground and isomeric states, the visible differences in their intensities on the TOF spectra should not appear; however, in our measurements, we find otherwise. Consider ${}^{97}{\rm Nb}$ and ${}^{97}{\rm Y}$, as shown in Fig.~\ref{Hist_TOF}, where the isomeric yield ratios vary considerably between 1+ and 2+. Given this fact, the isomeric yield ratio measurements were carried out for several nuclides across the 1+, 2+, and 3+. Figure~\ref{Hist} shows the observed TOF spectra for the various isomer-exhibiting nuclides studied; the horizontal axis has been converted to the mass-energy relative to the ground state of each nuclide. Contrary to expectation, the relative intensities of the isomeric states relative to the ground state were not constant with respect to the charge state. 

The nuclear spin difference ($\Delta I = I_{\rm isomer} - I_{\rm ground}$) dependence on the isomeric yield ratios can be found in the left panel of Fig.~\ref{esOfIsomer}. The $\Delta I$ ranges from 4 to 13 in the present measurements based on the spin values as evaluated in NUBASE2020 \citep{Kondev2021}. The yield ratios vary by four orders of magnitude, with a dependence on $\Delta I$ clearly visible, especially for the 1+. The ratios in the 2+ are distributed around unity with a weak $\Delta I $ dependence.

The nuclear spin effect governing the specific chemical reactions is known as the magnetic isotope effect (MagIE) \citep{Buchachenko2013} in chemistry, but this does not work in the present cases. The difference in chemical properties among different isotopes is called the isotope effect. This originates from two properties of atomic nuclei. The influence of differences in isotopic masses is categorized as the mass-dependent isotope effect (MDE) and is explained by the Bigeleisen-Myer theory \citep{Bigeleisen1974}. In contrast, the size and shape of atomic nuclei result in the nuclear field shift effect (NFSE) \citep{Yang2016}, which is known as one of the mass-independent isotope effects (MIEs). The MagIE is a kind of MIE, based on the spin conversion of a two-electron system in the pairs of radicals, which are atoms and molecules with unpaired valence electrons. Neutral helium atoms have coupled two valence electrons, and the total electron spin is always zero. Thus, there is no room for the MagIE to work if the contribution of the ETs with the neutral helium atoms is dominant in determining the ICSDs.

A comparison of the observed ICSDs with the MDE and the NFSE will be considered. To quantify the change of the ion yield ratio associated with the charge state changing from $q$ to $q^{\prime}$, in an analogy to the isotope enrichment factor \citep{Bigeleisen1974}, we introduce a spin enrichment factor, $\varepsilon_{\rm S}$, defined to be
\begin{equation}
\varepsilon_{\rm S}(q, q^{\prime}) \equiv \frac{\rho(q^{\prime})}{\rho(q)} - 1,
\label{DefineOfES}
\end{equation}
where $\rho$ is an ion yield ratio between the higher- and lower-spin states, represented by
\begin{equation}
\rho(q) \equiv \frac{Y_{\rm High}(q)}{Y_{\rm Low}(q)}.
\label{DefineOfRho}
\end{equation}
With this definition, a positive (negative) $\varepsilon_{\rm S}$-value means that the higher (lower) spin component is concentrated along with the change in the charge state.

The obtained ground and isomeric state pairs' $\varepsilon_{\rm S}$, plotted in the right panel of Fig.~\ref{esOfIsomer}, cannot be explained with the framework of any known isotope effects. The same measurements were performed several times for some nuclides to confirm reproducibility. They are indicated with different symbols. We observe no case that both $\varepsilon_{\rm S} (3+,2+)$ (hereafter denoted as $\varepsilon_{\rm S,32}$) and $\varepsilon_{\rm S,21}$ are zero within $1 \sigma$, indicating that this ICSD anomaly is reproducible. The absolute values of $\varepsilon_{S}$ are an order of $10^{-1}$ and are sufficiently large compared with the typical isotope enrichment factor (ranging from $10^{-4}$ to $10^{-3}$) due to the MDE and the NFSE.  No definitive correlation to a mass number is confirmed for the yttrium and niobium isotopes. Thus, the observed ICSD does not match the known MDE and MIE features. 

The CHeGC condition affected the yield ratios, but its influence on them also depends on the isotopes. In each measurement series, it was slightly different; in the first series, the intensities of impurities were higher than all others, as shown by the 1+ TOF spectra in Fig.~\ref{Hist_TOF}. The $\varepsilon_{\rm S,32}$ values of ${}^{97}{\rm Y}{}^{\rm m1,m2}$ at the first series (open circles in Fig.~\ref{esOfIsomer}) seem to be outliers. In contrast, the ${}^{98}{\rm Y}$ case shows consistent results in all three series. This implies that the sensitivity to the impurities in the filling helium gas can vary for different isotopes and nuclear states. \\

If the difference in the spin values of the ground and isomeric states causes an ICSD anomaly, non-zero $\varepsilon_{\rm S}$ should be observed for the pairs of isotopes for given elements with different ground state spins. To investigate this, we performed similar measurements, measuring the relative yields across the 1+, 2+, and 3+ along isotopic chains for several elements that could be extracted from the CHeGC.

The results for several such isotope pairs are plotted in Fig.~\ref{esOfIsotope}, and most cases indeed show the non-zero $\varepsilon_{\rm S}$ values, even if the even-$A$ -- even-$A$ pairs of the even-$Z$ element. This is also surprising because the ground state spin of even-$Z$ -- even-$A$ isotope is always zero. There is no hyperfine interaction for zero-spin atomic nuclei, and the electrons' states are the same in these even-$A$ isotopes (the known MDE and MIE remain, but their magnitude is negligible in the present measurements).

We point out that the shape difference in atomic nuclei might be related to these non-zero $\varepsilon_{\rm S}$-values, which cannot be explained solely by nuclear spin. The drop of $\varepsilon_{S}$ trend can be found between $A = 99$ and $A = 101$ in the zirconium case. The result of the zirconium isotope laser spectroscopy measurement \citep{Thayer2003} as indicated in the bottom plot of Fig~\ref{esOfIsotope} shows a change of the trend of $\delta \langle r^2 \rangle$ at $A = 100$, meaning a nuclear shape transition exists. For the tellurium isotopes, the $\delta \langle r^2 \rangle$ trend varies at $A = 135$ \citep{Sifi2007}, and the smooth increase of the nuclear deformation is confirmed for the cerium isotopes of $A = 140$ to 148 \citep{Cheal2003}. These experimental results are consistent with the observed $\varepsilon_{\rm S}$ trends of both elements, but further discussion is difficult for now due to the lack of experimental data regarding $\delta \langle r^2 \rangle$. 

Finally, it is worth noting that one possible reason this anomaly had not been observed previously may be the low abundance -- typically less than 10\%, with few exceptions -- of isotopes with a finite nuclear spin in the elements heavier than iron. Thus, investigating this phenomenon becomes challenging when only stable nuclides are considered.

We have discovered an ion charge state distribution anomaly, indicating that the state of the atomic nuclei affects the electron transfer between helium atoms and the radioisotope ions having various nuclear states. The framework of the mass-dependent and mass-independent isotope effects cannot explain this anomaly. The mechanisms of this phenomenon remain an open question \citep{EndMat}. The fundamental nuclear data needed to uncover this is insufficient yet. \\

\textit{Acknowledgments}--This work was supported by the Japan Society for the Promotion of Science, KAKENHI, Grant Number 17H06090, 22H04946, 19K14750, and 23K13137.


\bibliography{Isomer_charge_state_anomaly}

\section{End Matter}

\textit{A possible mechanism that explains an ICSD anomaly}--First, let us consider the case of the isomeric yield ratios. For most of the yttrium and niobium isotopes and the antimony isotopes, $\varepsilon_{S}$ satisfies the magnitude relationship:
\begin{equation}
 | \varepsilon_{\rm S,21} | > |\varepsilon_{\rm S,32} | ~~{\rm and}~~ \varepsilon_{\rm S,21} <  \varepsilon_{\rm S,32}. 
 \label{RelateOfES}
\end{equation}
The first relation, $\left| \varepsilon_{\rm S,21} \right| > \left| \varepsilon_{\rm S,32} \right|$, would originate from the fact that the abundance of the 2+ state is dominant in the ICSD, while the 1+ state is a minor term. Because a small change in the yield ratio of the 2+ ions significantly impacts the yield ratio of the 1+ ions. The simplest supposition leading to the second relation, $\varepsilon_{\rm S,21} < \varepsilon_{\rm S,32}$, is that only the lower-spin components of 2+ ions go to the 1+: this process results in $\varepsilon_{\rm S,32} < 0$ and $\varepsilon_{\rm S,32} > 0$. However, the negative $\varepsilon_{\rm S,32}$-values are also observed in the right panel of Fig.~\ref{esOfIsomer}, which means the influx of the lower-spin components from the 3+ contributes. Thus, the trend of Eq.~\ref{RelateOfES} implies the existence of ``relative'' enhancement of the electron capture process of the lower-spin components, which is a major point from the observed ICSDs.

An intermediate quasi-molecular state (QMS) in the ET process should be the key to understanding this anomaly. For that, the ET under the condition where the charge-exchange equilibrium is frozen will be considered in the following. We start by setting two assumptions. The first one is that the collision of incoming ions always occurs against neutral helium atoms. A contribution from possible impurities is neglected with this. Such an assumption seems justified by the relative lack of observed radio-molecular ions.  The second assumption is the low-energy limit; the kinetic energy of the incoming ion, M, is smaller than around $0.02~{\rm MeV/nucleon}$, where the relative velocity to a helium atom is slower than the Bohr velocity. In this condition, the collision time is sufficiently longer for a valence electron to go around the colliding atoms. Consequently, the electron capture reaction occurs via the forming of a quasi-molecular state (QMS)
\begin{equation}
{\rm M}^{(n+1)+} + {\rm He} \rightarrow ({\rm MHe})^{\ast (n+1)+} \rightarrow {\rm M}^{n+} + {\rm He}^{+},
\label{eCapt}
\end{equation}
only when the energy relation between their ionization potentials ($IP$s) and the external energy contribution ($E_{\rm exter}$) satisfies the relation of $IP_{n+1}({\rm M})+E_{\rm exter} > IP_{1}( {\rm He}) = 24.6~{\rm eV}$. In the inverse - the electron stripping reaction of the ions - two different processes can be considered:
\begin{eqnarray}
\label{eStrip1}
{\rm M}^{(n-1)+} \hspace{-0.2em} + \hspace{-0.2em} {\rm He}  \hspace{-0.2em} \rightarrow \hspace{-0.2em} {\rm M}^{\ast (n-1)+} \hspace{-0.2em} +\hspace{-0.2em} {\rm He} \hspace{-0.2em} \rightarrow {\rm M}^{n+}  \hspace{-0.2em} + \hspace{-0.2em} {\rm e}^{-}  \hspace{-0.2em} + \hspace{-0.2em} {\rm He},&& \\
\label{eStrip2}
{\rm M}^{(n-1)+} + {\rm He} \rightarrow ({\rm MHe})^{\ast (n-1)} \rightarrow {\rm M}^{n+} +{\rm He}^{\ast -}.&&
\end{eqnarray}
For Eq.~\ref{eStrip1}, an external energy contribution exceeding the $IP_{\rm n}({\rm M})$ is needed to occur. Forming a helium anion, ${\rm He}^{\ast -}$, from ground-state helium requires an additional energy of $19.8~{\rm eV}$ \citep{Mauracher2014}. Thus,  $IP_{\rm n}({\rm M})+19.8~{\rm eV}$ is the threshold of the process of Eq.~\ref{eStrip2}. From the experimental results regarding the ion equilibrium charge state in helium gas \citep{Gregorich2013}, the external energy contribution is insufficient to keep the charge-exchange equilibrium in the present low-energy condition. This indicates that once ICSD deviates from that of the equilibrium, it is difficult to return to it.

In the low-energy limit, the instantaneous ET described by the velocity matching between the helium's valence electron and the unoccupied orbits of the ions no longer occurs. Then, forming the QMS is crucial in the ET process. Nuclear spin could prevent the forming of the QMS if it significantly changes the effective potential or the configuration of the molecular orbits of the QMS, which is occupied by a transferring electron. This could suppress the ET processes described by Eqs.~\ref{eCapt} and \ref{eStrip2} for the higher-spin case, while the electron stripping process of Eq.~\ref{eStrip1} can still occur. Thus, the fraction of the lower charge state, such as the 1+, in the ISCD of the higher-spin components decreases. In contrast, all three ET processes would be allowed in the lower-spin case. Consequently, the relative enhancement of the electron capture process of the lower-spin components would be established, and the ICSD differences would appear depending on nuclear spin. 

As discussed above, the results of Fig~\ref{esOfIsotope} cannot be explained by considering only the nuclear spin difference, and the relation to the nuclear deformation is pointed out. The possible mechanism is as follows: if deformed, non-spherical Coulomb potential caused by a highly deformed atomic nucleus could change the configuration of the QMS, the nuclear shape difference would be one of the possible origins of an ICSD anomaly. The influences originating from nuclear spin and deformation can coexist with this hypothesis.

\begin{figure}[t]
	\begin{center}
	        \includegraphics[width=0.5\textwidth, bb=0 0 842 595, clip, trim=0 250 0 0]{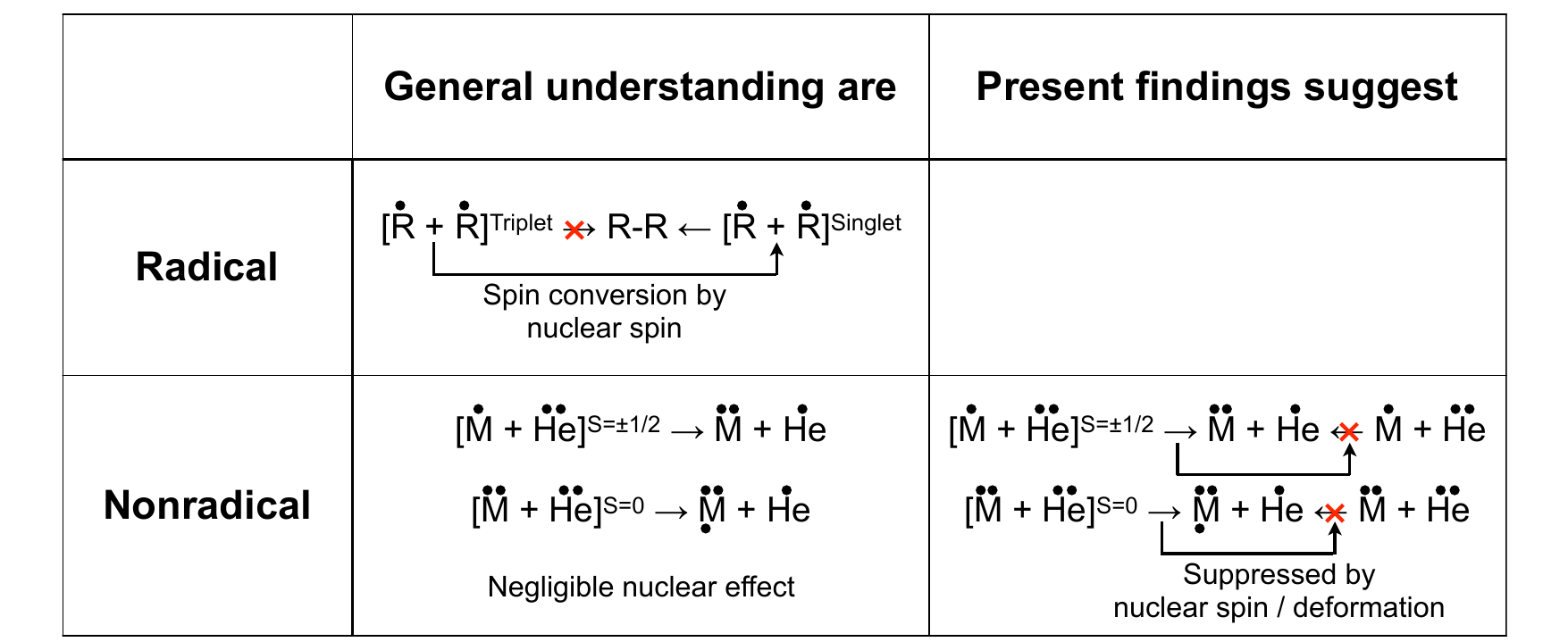}
		\caption{Summary of the proposed mechanism of an ICSD anomaly. The dots on each character represent valence electrons. The electron states, or total spin values, are indicated. \label{Summary}}
	\end{center}
\end{figure}

Our proposed mechanism, which can explain an ICSD anomaly, is summarized in Fig~\ref{Summary}. The two different types of electron systems are shown. The first is the pair of radicals, where the MagIE governs the reaction. The second is the case in which the MagIE cannot work within the general understanding: the reaction system contains the nonradical(s) as at least one of the reactants. If the possible mechanism is reasonable, the present findings suggest that nuclear spin and deformation change the reaction characteristics even if the system is intuitively insensitive to the nuclear effects.

\section{Supplemental Material}

\subsection{Method}

The peak identification was done by determining ion mass. The mass-determining way is similar to the previous works with the MRTOF-MS \citep{Kimura2024}. The measured TOF value for an ion, having mass $m$ and charge $q$,  which undergoes $n$ laps in the MRTOF-MS device, can be represented by 
\begin{equation}
t_{\rm obs} = (a + b \cdot n) \sqrt{m/q} + t_0
\label{eqObsTOF}
\end{equation}
where $a$ and $b$ are constants related to the non-reflection flight path and the path between consecutive reflections, respectively, and $t_0$ represents an electronic delay between the TDC start signal and the ion's actual ejection from the flat trap. The single reference method \citep{Ito2013}, which needs only one reference mass, is adopted to determine the mass of analyte ions. In this method $m_{\rm X}$, the ionic mass of nuclide X, is given by Eq.~\ref{eqSingleRef}:
\begin{equation}
m_{\rm X} = \rho_{\rm tof}^2 m_{\rm ref} = \left( \frac{t_{\rm X}-t_0}{t_{\rm ref}-t_0} \right)^2 m_{\rm ref},
\label{eqSingleRef}
\end{equation}
where $\rho_{\rm tof}$ is the TOF ratio, $t_{\rm X}$ and $t_{\rm ref}$ are the TOF of nuclide X and the reference ion, respectively, $m_{\rm ref}$ is the mass of the reference ion,  $t_0$ is the constant time offset within the measurement system mentioned above. 

A phenomenological fitting function, based on an exponential-Gaussian hybrid function \citep{Lan2001}, was used to fit non-Gaussian-shape peaks accurately. In the present study, we employed the function:
\begin{align}
f(\tau) = \left\{
\begin{array}{l}
\left( \kappa / \sigma \right) \exp \left[ \frac{t_{\rm s1} (t_{\rm s1} -2\tau)}{2\sigma^2} \right] 
\ \left({\rm for} \ \tau < t_{\rm s1}\right), \\
\\
\left( \kappa / \sigma \right) \exp \left[ - \frac{\tau^2}{2\sigma^2} \right] 
\ \left({\rm for} \ t_{\rm s1} \leq \tau < t_{\rm s2}\right)\\
\\
\left( \kappa / \sigma \right) \exp \left[ \frac{t_{\rm s2} (t_{\rm s2} -2\tau)}{2\sigma^2} \right] 
\ \left({\rm for} \ \tau \geq t_{\rm s2}\right),
\end{array}
\right.
\end{align}
where $t_{{\rm s}i}$ denotes  the range of each sub-function. The variable $\tau$ is defined as $\tau \equiv t - \mu$, where $\mu$ is the peak center used in the mass determinations. 

Herein, we set a presumption about the peak shape: the peaks of the ions belonging to the same $A/q$ series have an identical shape. Based on this assumption, the only free parameters for each peak in the fitting function are the peak center $\mu$ and the peak height $\kappa$ for the species of interest. In the fitting algorithm, to improve the mass precision, the $\tau$ parameter was treated as a function of $t$, $t_{\rm ref}$, and $\rho_{\rm tof}$. Then, the fitting function $F$ for $N$ peaks was described by 
 \begin{eqnarray}
F(t, t_{\rm ref}, \rho_{\rm tof~1}, \cdots, \rho_{\rm tof~N},, \kappa_1, \cdots, \kappa_{\rm N}) =  \nonumber  \\
\sum_{i=1}^{N} f(t, t_{\rm ref}, \rho_{{\rm tof}~i},\kappa_i, \sigma, t_{{\rm s}1}, t_{{\rm s}2}).
\end{eqnarray}

The ion yield ratios were determined using peak fit and event counting. The former was used for the ground and isomeric state pairs. In this case, the $\rho(q)$ can be represented by the peak hight parameter $\kappa$ as follows;
\begin{equation} 
\rho(q) = \frac{\kappa_{\rm High}}{\kappa_{\rm Low}}.
\end{equation}
Event counting was applied to the isotope pair measurements because the presumption regarding the peak shape is not always valid. An integrated region for event counting was determined by the peak fit to a single $A/q$ series, and then they were calculated. In the ideal condition that two peaks are separated and their statistics are high enough, both methods must return the same value. For example, in the case of ${}^{96}{\rm Y}{}^{2+}$, $\rho(2+) = 0.790(14)$ is obtained via the event counting and agrees on the value of the peak fit of 0.791(51).

\subsection{Tables}

The experimental condition of each series of measurements is given in Table~\ref{MeasCond}. The raw measured values of $\rho$ with the spin-parity information on the corresponding nuclear states of interest are tabulated in Tables~\ref{Summary_Isomer} and \ref{Summary_Isotope}. 

\begin{table*}[h]
\caption{Information on the measurement conditions. A single helium cylinder was used for each series of measurements. \label{MeasCond}}
\begin{tabular}{ccccc}
\hline \hline
\multirow{2}{*}{Meas. \#} & \multirow{2}{*}{Date} & \multicolumn{2}{c}{${}^{252}$Cf source}                 & \multirow{2}{*}{Gas-cell temp.} \\
                          &                       & \multicolumn{1}{c}{Serial \#} & Ref. date     &                          \\
\hline
1                         & March 2023            & S4-608                        & 15th March 2020 & 47.1 K                     \\
2                         & December 2023            & S4-608                        & 15th March 2020 & 55.9 K                     \\
3                         & January 2024             & W6-031                        & 1st December 2023 & 56.0 K                     \\
4                         & February 2024             & W6-031                        & 1st December 2023 & 46.6 K                     \\
\hline \hline 
\end{tabular}
\end{table*}

\begin{table*}[h]
\caption{Observed intensity ratio between ground and isomer states and the related nuclear properties. All values of $J^{\pi}$ are taken from NUBASE2020 and the brackets indicate that the corresponding values are predicted ones.\label{Summary_Isomer}}
\renewcommand{\arraystretch}{1.25}
\begin{tabular}{cccclll}
\hline \hline
              & \multicolumn{2}{c}{$J^{\pi}$} & &\multicolumn{3}{c}{$\rho$} \\
 Nuclide &  ground state & isomeric state & Meas. \# &\multicolumn{1}{c}{1+} &  \multicolumn{1}{c}{2+}  &  \multicolumn{1}{c}{3+} \\
\hline
${}^{93}$Y     & 1/2-  & 9/2+        & 1 & 0.199(30)              & 0.497(58)    & 1.07(12) \\
\hline
${}^{96}$Y     & 0-     & 8+           & 1 & 0.229(27)              & 0.791(51)    & 0.667(19) \\
\hline
${}^{97}$Y,m1  & 1/2-  & 9/2+      & 1 & 0.73(10)                & 2.16(11)      & 1.682(92) \\
                        &          &              & 3 & 0.279(82)              & 2.66(12)      & 2.82(19) \\
                        &          &              & 4 & 0.292(15)              & 2.302(67)     & 2.895(35) \\
\hline
${}^{97}$Y,m2  & 1/2-  & (27/2-)  & 1 & 0.0018$^{\dag}$   & 0.0462(30)  & 0.0284(22) \\
                        &          &              & 3 & 0.0034$^{\dag}$   & 0.1212(68)      & 0.142(10) \\
                        &          &              & 4 & 0.0017$^{\dag}$   & 0.1136(60)     & 0.1412(40) \\
\hline
${}^{98}$Y       & 0-     & (6,7)+  & 1 & 0.0108(18)           & 0.1698(83)    & 0.274(17) \\
                        &          &              & 3 & 0.00674(93)         & 0.1383(51)    & 0.196(14) \\
                        &          &              & 4 & 0.0199(26)           & 0.1027(28)    & 0.2119(60) \\
 \hline
${}^{97}$Nb   & 9/2+ & 1/2-         & 1 & 0.017$^{\dag}$     & 12.5(62)      & 8.2(25) \\
                        &          &              & 4 & 0.065$^{\dag}$    &  12.4(34)     & 8.4(18) \\
\hline
${}^{99}$Nb   & 9/2+ & 1/2-         & 1 & 1.93(17)                & 2.74(17)      & 2.59(18) \\
                        &          &              & 3 & 2.080(54)             & 2.957(87)    & 3.02(17) \\
                        &          &              & 4 & 1.930(63)             & 2.499(66)    & 3.016(79) \\
 \hline
${}^{100}$Nb   & 1+   & (5+)        & 4 & 1.745(40)                & 0.830(20)      & 2.763(63) \\
\hline
${}^{129}$Sb & 7/2+ & 19/2-     & 1 & 0.333(20)              & 0.408(20)    & 0.379(63) \\
                        &          &              & 4 & 0.339(12)              & 0.4306(98)    & 0.366(38) \\
 \hline
${}^{133}$Te &  (3/2+) & (11/2-)   & 4 & 2.496(49)              & 2.996(31)    & 9.91(88) \\
\hline \hline
\end{tabular}
\begin{tablenotes}
\item[1] \hspace{37.5em} ${}^{\dag}$ Upper limit.
\end{tablenotes}
\end{table*}

\begin{table*}[h]
\caption{Observed intensity ratio between the isotopes with different mass numbers. All values of $J^{\pi}$ are taken from NUBASE2020 and the brackets indicate that the corresponding values are predicted ones.\label{Summary_Isotope}}
\renewcommand{\arraystretch}{1.25}
\begin{tabular}{cccccclll}
\hline \hline
              & \multicolumn{2}{c}{Lower state} & \multicolumn{2}{c}{Higher state} &&\multicolumn{3}{c}{$\rho$} \\
 Element &  $A$ & $J^{\pi}$  & $A$ & $J^{\pi}$ & Meas. \# &\multicolumn{1}{c}{1+} &  \multicolumn{1}{c}{2+}  &  \multicolumn{1}{c}{3+} \\
\hline
 \multicolumn{2}{l}{Even atomic number} &&&&&&&\\
Zr   &  100  & 0+  &  98  & 0+         & 2 & 2.53(12)              & 0.460(13)    & 0.258(20) \\
      &  100  & 0+  &  99  & 1/2+    & 2 & 2.797(77)            & 0.655(14)    & 0.5547(94) \\
      &  100  & 0+  &  101  & 3/2+   & 2 & 0.471(24)           & 0.797(12)    & 0.843(13) \\
      &  100  & 0+  &  102  & 0+        & 2 & 0.536(21)            & 0.695(15)    & 0.671(11) \\
\hline
Mo &  106  & 0+  &  105  & (5/2-)    & 2 & 1.231(20)              & 1.0248(65)    & 1.056(14) \\
      &  106  & 0+  &  107  & (1/2+)   & 2 & 2.547(37)            & 0.4530(37)    & 0.3614(63) \\
      &  106  & 0+  &  108  & 0+         & 2 & 0.2030(60)           & 0.1997(46)    & 0.1910(91) \\
 \hline
Te  &  134  & 0+  &  132  & 0+                  & 2 & 0.3004(33)                 & 0.439(18)                 & 0.522(31) \\
      &  134  & 0+  &  133  & (3/2+),(11/2-)  & 2 & 0.7749(92)$^{\dag}$  & 0.596(17)$^{\dag}$  & 0.2388(41)$^{\dag}$ \\
      &  134  & 0+  &  135  & (7/2-)             & 2 & 0.5903(76)                 & 0.620(18)                 & 0.641(17) \\
      &  134  & 0+  &  136  & 0+                  & 2 & 0.3745(56)                 & 0.2484(99)               & 0.0307(14) \\
      &  134  & 0+  &  138  & 0+                  & 2 & 0.0346(10)                 & 0.159(10)                 & 0.0095(34) \\
\hline
Ce  &  148  & 0+  &  146  & 0+       & 2 & 1.820(41)           & 0.499(14)      & 0.560(13) \\
      &  148  & 0+  &  147  & (5/2-)   & 2 & 1.139(23)           & 0.728(12)      & 0.7240(83) \\
      &  148  & 0+  &  149  & (3/2-)   & 2 & 0.415(11)            & 0.798(11)      & 0.7746(87) \\
      &  148  & 0+  &  150  & 0+        & 2 & 0.1708(64)         & 0.5615(99)    & 0.5784(71) \\
\hline
Nd &  152  & 0+  &  151  & 3/2+     & 2 & 1.466(49)         & 0.488(11)    & 0.445(19) \\
      &  152  & 0+  &  153  & (3/2)-   & 2 & 1.93(11)            & 1.016(22)    & 0.963(34) \\
      &  152  & 0+  &  154  & 0+        & 2 & 0.876(34)         & 0.961(18)    & 0.904(31) \\
\hline
 \multicolumn{2}{l}{Odd atomic number} &&&&&&&\\
La  &  149  & (3/2-)  &  147  & (5/2+)   & 2 & 0.314(14)            & 2.653(84)      & 5.828(96) \\
      &  149  & (3/2-)  &  148  & (2-)       & 2 & 0.736(25)            & 2.099(58)      & 2.361(43) \\
      &  149  & (3/2-)  &  150  & (3+)      & 2 & 0.0106(23)          & 0.0831(81)    & 0.2740(90) \\
\hline
Pr  &  150  & 1-  &  149  & (5/2+)    & 2 & 1.148(27)             & 0.8023(78)    & 0.847(23) \\
      &  150  & 1-  &  151  & (3/2-)     & 2 & 2.785(57)            & 1.090(14)       & 1.041(27) \\
      &  150  & 1-  &  152  & 4+       & 2 & 0.427(14)            & 0.6774(69)     & 0.807(22) \\
\hline
Pm &  155  & (5/2-)  &  153  & 5/2-          & 2 & 1.687(67)                  & 0.3716(63)                 & 0.254(24) \\
      &  155  & (5/2-)  &  156  & 4+,(1+)   & 2 & 0.707(35)$^{\dag}$   & 0.834(11)$^{\dag}$    & 1.102(64)$^{\dag}$ \\
\hline \hline
\end{tabular}
\begin{tablenotes}
\item[1]  \hspace{23em} ${}^{\dag}$ The ratio to the sum of ground and isomeric states.
\end{tablenotes}
\end{table*} 

\end{document}